\documentclass[preprint]{elsarticle}
\usepackage{geometry}
\usepackage{graphicx}
\usepackage[caption=false,font=footnotesize]{subfig}
\usepackage{float}
\usepackage{fixltx2e}
\usepackage{longtable}

\usepackage{comment}
\usepackage{amsmath}
\usepackage{amssymb}
\usepackage{xtab}
\usepackage{multirow}
\usepackage{url}
\usepackage[numbers]{natbib}

\begin{document}
\title{Testing Scientific Software: A Systematic Literature Review}
\author{Upulee Kanewala\corref{cor1}}
\ead{upuleegk@cs.colostate.edu}
\author{James M. Bieman}
\ead{bieman@cs.colostate.edu}

\cortext[cor1]{Corresponding author}

\address{Computer Science Department, Colorado State University, USA}

\begin{abstract}

Context: Scientific software plays an important role in critical decision making, for example making weather predictions based on climate models, and computation of evidence for research publications. Recently, scientists have had to retract publications due to errors caused by software faults. Systematic testing can identify such faults in code. 

Objective: This study aims to identify specific challenges, proposed solutions, and unsolved problems faced when testing scientific software. 

Method: We conducted a systematic literature survey to identify and analyze relevant literature. We identified 62 studies that provided relevant information about testing scientific software. 

Results: We found that challenges faced when testing scientific software fall into two main categories: (1) testing challenges that occur due to characteristics of scientific software such as oracle problems and (2) testing challenges that occur due to cultural differences between scientists and the software engineering community such as viewing the code and the model that it implements as inseparable entities. In addition, we identified methods to potentially overcome these challenges and their limitations. Finally we describe unsolved challenges and how software engineering researchers and practitioners can help to overcome them. 

Conclusions: Scientific software presents special challenges for testing. Specifically, cultural differences between scientist developers and software engineers, along with the characteristics of the scientific software make testing more difficult. Existing techniques such as code clone detection can help to improve the testing process. Software engineers should consider special challenges posed by scientific software such as oracle problems when developing testing techniques.

\end{abstract}

\begin{keyword}
Scientific software \sep Software testing \sep Systematic literature review \sep Software quality
\end{keyword}
\maketitle
\section{Introduction}
Scientific software is widely used in science and engineering fields. Such software plays an important role in critical decision making in fields such as the nuclear industry, medicine and the military~\cite{sanders08challenge,4548404}. For example, in nuclear weapons simulations, code is used to determine the impact of modifications, since these weapons cannot be field tested~\cite{Post:2004:SPM}. Climate models make climate predictions and assess climate change~\cite{Drake:2005:OSD:1093627.1093634}. In addition, results from scientific software are used as evidence in research publications~\cite{4548404}. Due to the complexity of scientific software and the required specialized domain knowledge, scientists often develop these programs themselves or are closely involved with the development~\cite{gmd-5-1009-2012,chinASurvey2007,SegalModels2008,Carver:2007:SDE:1248820.1248886}. But scientist developers may not be familiar with accepted software engineering practices~\cite{
SegalModels2008,sanders08challenge}. This lack of familiarity can impact the quality of scientific software~\cite{Easterbrook:2010:CCG:1882362.1882383}. 

Software testing is one activity that is impacted. Due to the lack of systematic testing of scientific software, subtle faults can remain undetected. These subtle faults can cause program output to change without causing the program to crash. Software faults such as one-off errors have caused the loss of precision in seismic data processing programs~\cite{609829}. Software faults have compromised coordinate measuring machine (CMM) performance~\cite{ABACKERLI2010}. In addition, scientists have been forced to retract published work due to software faults~\cite{Miller22122006}. Hatton \textit{et al.} found that several software systems written for geoscientists produced reasonable yet essentially different results~\cite{328993}. There are reports of scientists who believed that they needed to modify the physics model or develop new algorithms, but later discovered that the real problems were small faults in the code~\cite{6241365}.     

We define scientific software broadly as \emph{software used for scientific purposes}. Scientific software is mainly developed to better understand or make predictions about real world processes. The size of this software ranges from 1,000 to 100,000 lines of code~\cite{4548404}. Developers of scientific software range from scientists who do not possess any software engineering knowledge to experienced professional software developers with considerable software engineering knowledge. 

To develop scientific software, scientists first develop discretized models. These discretized models are then translated into algorithms that are then coded using a programming language. Faults can be introduced during all of these phases~\cite{Dahlgren:2005:ISS:1145319.1145341}. Developers of scientific software usually perform validation to ensure that the scientific model is correctly modeling the physical phenomena of interest~\cite{5228715,Murphy:2011:ETH:1987993.1988003}. They perform verification to ensure that the computational model is working correctly~\cite{5228715}, using primarily mathematical analyses~\cite{Post:2004:SPM}. But scientific software developers rarely perform systematic testing to identify faults in the code~\cite{KellyAssesing2008,Murphy:2011:ETH:1987993.1988003,5069163,sanders08challenge}. Farrell \textit{et al.} show the importance of performing code verification to identify differences 
between the code and the discretized model~\cite{gmd-4-435-2011}. Kane \textit{et al.} found that automated testing is fairly uncommon in biomedical software development~\cite{DBLP:journals/bmcbi/KaneHCMKB06}. In addition, Reupke \textit{et al.} discovered that many of the problems found in operational medical systems are due to inadequate testing~\cite{5439}. Sometimes this lack of systematic testing is caused by special testing challenges posed by this software~\cite{Easterbrook:2010:CCG:1882362.1882383}. 

This work reports on a Systematic Literature Review (SLR) that identifies the special challenges posed by scientific software and proposes solutions to overcome these challenges. In addition, we identify unsolved problems related to testing scientific software. 

An SLR is a ``means of evaluating and interpreting all available research relevant to a particular research question or topic area or phenomenon of interest''~\cite{citeulike:1191541}. The goal of performing an SLR is to methodically review and gather research results for a specific research question and aid in developing evidence-based guidelines for the practitioners~\cite{Kitchenham20097}. Due to the systematic approach followed when performing an SLR, the researcher can be confident that she has located the required information as much as possible.

Software engineering researchers have conducted SLRs in a variety of software engineering areas. Walia \textit{et al.}~\cite{Walia20091087} conducted an SLR to identify and classify software requirement errors. Engstr\"{o}m \textit{et al.}~\cite{Engström201014} conducted an SLR on empirical evaluations of regression test selection techniques with the goal of ``finding a basis for further research in a joint industry-academia research project''. Afzal \textit{et al.}~\cite{Afzal2009957} carried out an SLR on applying search-based testing for performing non-functional testing. Their goal is to ``examine existing work into non-functional search-based software testing''. While these SLRs are not restricted to software in a specific domain, we focus on scientific software, an area that has received less attention than application software. Further when compared to Engstr\"{o}m \textit{et al.} or Afzal \textit{et al.}, we do not restrict our SLR to a specific testing technique.  

The overall goal~\cite{Kitchenham20097} of our SLR is to \emph{identify specific challenges faced when testing scientific software, how the challenges have been met, and any unsolved challenges.} We developed a set of research questions based on this overall goal to guide the SLR process. Then we performed an extensive search to identify publications that can help to answer these research questions. Finally, we synthesized the gathered information from the selected studies to provide answers to our research questions.  

This SLR identifies two categories of challenges in scientific software testing. The first category are challenges that are due to the characteristics of the software itself such as the lack of an oracle. The second category are challenges that occur because scientific software is developed by scientists and/or scientists play leading roles in scientific software development projects, unlike application software development where software engineers play leading roles. We identify techniques used to test scientific software including techniques that can help to overcome oracle problems and test case creation/selection challenges. In addition, we describe the limitations of these techniques and open problems.

This paper is organized as follows: Section~\ref{sec:researchMethod} describes the SLR process and how we apply it to find answer to our research questions. We report the findings of the SLR in Section~\ref{sec:reviewReport}. Section~\ref{sec:discussion} contains the discussion on the findings. Finally we provide conclusions and describe future work in Section~\ref{sec:conc}.

\section{Research Method}
\label{sec:researchMethod}

We conducted our SLR following the published guidelines by Kitchenham~\cite{citeulike:1191541}. The activities performed during an SLR can be divided into three main phases: (1) planning the SLR, (2) conducting the review and (3) reporting the review. We describe the tasks performed in each phase below.


%

\subsection{Planning the SLR}
\subsubsection{Research Questions}
\label{sec:RQs}
 
The main goal of this SLR is to identify specific challenges faced when testing scientific software, how the challenges have been met, and any unsolved challenges. We developed the following research questions to achieve our high level goal:  
\begin{enumerate}
 \item[\textbf{RQ1:}] How is scientific software defined in the literature?
 \item[\textbf{RQ2:}] Are there special characteristics or faults in scientific software or its development that make testing difficult?
 \item[\textbf{RQ3:}] Can we use existing testing methods (or adapt them) to test scientific software effectively?
 \item[\textbf{RQ4:}] Are there challenges that could not be met by existing techniques?
\end{enumerate}

\subsubsection{Formulation and validation of the review protocol}

The review protocol specifies the methods used to carry out the SLR. Defining the review protocol prior to conducting the SLR can reduce researcher bias~\cite{KitchenhamGuidlines2007}. In addition, our review protocol specifies source selection procedures, search process, quality assessment criteria and data extraction strategies.

\noindent\textbf{Source selection and search process:} We used the Google Scholar, IEEE Xplore, and ACM Digital Library databases since they include journals and conferences focusing on software testing as well as computational science and engineering. Further, these databases provide mechanisms to perform key word searches. We did not specify a fixed time frame when conducting the search. We conducted the search in January 2013. Therefore this SLR includes studies that were published before January 2013. We did not search for specific journals/conferences since an initial search found relevant studies published in journals such as Geoscientific Model Development\footnote{http://www.geoscientific-model-development.net/} that we were not previously familiar with. In addition, we examined relevant studies that were referenced by the selected primary studies.

We searched the three databases identified above using a search string that included the important key words in our four research questions. Further, we augmented the key words with their synonyms, producing the following search string:
\begin{quote}
(((challenges \textbf{OR} problems \textbf{OR} issues \textbf{OR} characteristics) \textbf{OR} (technique \textbf{OR} methods \textbf{OR} approaches)) \textbf{AND} (test \textbf{OR} examine)) \textbf{OR} (error \textbf{OR} fault \textbf{OR} defect \textbf{OR} mistake \textbf{OR} problem \textbf{OR} imperfection \textbf{OR} flaw \textbf{OR} failure)  \textbf{AND} (``(scientific \textbf{OR} numeric \textbf{OR} mathematical \textbf{OR} floating point) \textbf{AND} (Software \textbf{OR} application \textbf{OR} program \textbf{OR} project \textbf{OR} product)'')
\end{quote}

\noindent\textbf{Study selection procedure:} We systematically selected the primary studies by applying the following three steps.
\begin{enumerate}
 \item We examined the paper titles to remove studies that were clearly unrelated to our search focus.
 \item We reviewed the abstracts and key words in the remaining studies to select relevant studies. In some situations an abstract and keywords did not provide enough information to determine whether a study is relevant. In such situations, we reviewed the conclusions.
 \item We filtered the remaining studies by applying the inclusion/exclusion criteria given in Table~\ref{tab:IncExc}. Studies selected from this final step are the initial primary studies for the SLR.
\end{enumerate}
We examined the reference lists of the initial primary studies to identify additional studies that are relevant to our search focus.
\begin{table*}[!h]
  \centering
  \begin{tabular}{|p{0.475\textwidth}|p{0.475\textwidth}|}
  \hline
    \textbf{Inclusion criteria} & \textbf{Exclusion criteria}\\\hline
    \begin{enumerate}
     \item Papers that describe characteristics of scientific software that impact testing.
     \item Case studies or surveys of scientific software testing experiences.
     \item Papers that analyze characteristics of scientific software testing including case studies and experience reports.
     \item Papers describing commonly occurring faults in scientific software. 
      \item Papers that describe testing methods used for scientific software and provide a sufficient evaluation of the method used.
      \item Experience reports or case studies describing testing methods used for scientific software.
    \end{enumerate}
    &
    \begin{enumerate}
     \item Papers that present opinions without sufficient evidence supporting the opinion.
     \item Studies not related to the research questions.
     \item Studies in languages other than English.
     \item Papers presenting results without providing supporting evidence.
     \item Preliminary conference papers of included journal papers.
    \end{enumerate}\\
    \hline
  \end{tabular}
  \caption{Inclusion and exclusion criteria}
  \label{tab:IncExc}
\end{table*}

\noindent\textbf{Quality assessment checklist:} We evaluated the quality of the selected primary studies using selected items from the quality checklists provided by Kitchenham and Charters~\cite{KitchenhamGuidlines2007}. Table~\ref{tab:qualityForQuant} and Table~\ref{tab:qualityForQual} show the quality checklists that we used for quantitative and qualitative studies respectively. When creating the quality checklist for quantitative studies, we selected quality questions that would evaluate the four main stages of a quantitative study: design, conduct, analysis and conclusions ~\cite{KitchenhamGuidlines2007}.

\begin{table*}[!hbtp]
  \centering
  \begin{tabular}{|p{0.3\textwidth}|p{0.3\textwidth}|p{0.3\textwidth}|}
  \hline
  \textbf{Survey} & \textbf{Case study} & \textbf{Experiment}\\\hline
  \multicolumn{3}{|c|}{G1: Are the study aims clearly stated?}\\\hline
  S1: Was the method for collecting the sample data specified (e.g. postal, interview, web-based)?&N/A&N/A\\\hline
  S2: Is there a control group?&N/A&E1: Is there a control group?\\\hline
  N/A&N/A&E2: Were the treatments randomly allocated?\\\hline
  \multicolumn{3}{|c|}{G2: Are the data collection methods adequately described?}\\\hline
  \multicolumn{3}{|c|}{G3: Was there any statistical assessment of results?}\\\hline
  S3: Do the observations support the claims? &C1: Is there enough evidence provided to support the claims?&E3: Is there enough evidence provided to support the claims?\\\hline
  \multicolumn{3}{|c|}{G4: Are threats to validity and/or limitations reported?}\\\hline
  \multicolumn{3}{|c|}{G5: Can the study be replicated?}\\\hline
  \end{tabular}
  \caption{Quality assessment for quantitative studies}
  \label{tab:qualityForQuant}
\end{table*}

\begin{table}[!hbtp]
  \centering
  \begin{tabular}{|p{0.9\textwidth}|}
  \hline
  \textbf{Quality assessment questions}\\\hline
  \begin{enumerate}
   \item A: Are the study aims clearly stated?
   \item B: Does the evaluation address its stated aims and purpose?
   \item C: Is sample design/target selection of cases/documents defined?
    \item D: Is enough evidence provided to support the claims?
    \item E: Can the study be replicated?
  \end{enumerate}
  \\\hline
\end{tabular}
  \caption{Quality assessment for qualitative studies}
  \label{tab:qualityForQual}
\end{table}

\noindent\textbf{Data extraction strategy:} Relevant information for answering the research questions needed to be extracted from the selected primary studies. We used data extraction forms to make sure that this task was carried out in a accurate and consistent manner. Table~\ref{tab:dataForm} shows the data extraction from that we used.

\begin{table*}[!hbtp]
  \centering
  \begin{tabular}{|p{0.15\textwidth}|p{0.25\textwidth}|p{0.55\textwidth}|}
   \hline
  \textbf{Search focus}&\textbf{Data Item} &\textbf{Description}\\\hline
  \multirow{5}{*}{General}& Identifier & Reference number given to the article\\\cline{2-3}
  &Bibliography & Author, year, Title, source \\\cline{2-3}
  &Type of article & journal/conference/tech. report\\\cline{2-3}
  &Study aims & Aims or goals of the study\\\cline{2-3}
  &Study design & controlled experiment/survey/etc.\\\hline
  \multirow{2}{*}{RQ1}&Definition & Definition for scientific software \\\cline{2-3}
  &Examples &  Examples of scientific software\\\hline
  \multirow{3}{*}{RQ2}&Challenge/problem & Challenges/problems faced when testing scientific software\\\cline{2-3}
  &Fault description & Description of the fault found\\\cline{2-3}
  &Causes & What caused the fault?\\\hline
  \multirow{6}{*}{RQ3/RQ4}&Testing method & Description of the method used\\\cline{2-3}
  &Existing/new/extension & Whether the testing method is new, existing or modification to an existing method\\\cline{2-3}
  &Challenge/problem & The problem/challenge that it provides the answer to\\\cline{2-3}
  &Faults/failures found & Description of the faults/ failures found by the method\\\cline{2-3}
  &Evidence & Evidence for the effectiveness of the method of finding faults\\\cline{2-3}
  &Limitations & Limitations of the method\\\hline
  \end{tabular}
  \caption{Data extraction form}
  \label{tab:dataForm}
\end{table*}

\subsection{Conducting the review}

\subsubsection{Identifying relevant studies and primary studies}

The key word based search produced more than 6000 hits. We first examined paper titles to remove any studies that are not clearly related to the research focus. Then we used the abstract, key words and the conclusion to eliminate additional unrelated studies. After applying these two steps, 94 studies remained. We examined these 94 studies and applied the inclusion/exclusion criteria in Table~\ref{tab:IncExc} to select 49 papers as primary studies for this SLR.

Further, we applied the same selection steps to the reference lists of the selected 49 primary studies to find additional primary studies that are related to the research focus. We found 13 studies that are related to our research focus that were not already included in the initial set of primary studies. Thus, we used a total of 62 papers as primary studies for the SLR. The selected primary studies are listed in Tables~\ref{tab:primaryStudies1} and~~\ref{tab:primaryStudies2}. Table~\ref{tab:pubVenues} lists the publication venues of the selected primary papers. The International Workshop on Software Engineering for Computational Science and Engineering and the Journal of Computing in Science \& Engineering published the greatest number of primary studies.

\newcounter{rnum}
\setcounter{rnum}{0}
\begin{table*}
\footnotesize
\caption{Selected Primary Studies (Part 1)}
\label{tab:primaryStudies1}
\centering
\begin{tabular}{|l|l|p{0.55\textwidth}|c|c|c|c|}
\hline
\textbf{Study}&\textbf{Ref.}&\multirow{2}{*}{\textbf{Study focus}}&\multirow{2}{*}{\textbf{RQ1}}&\multirow{2}{*}{\textbf{RQ2}}&\multirow{2}{*}{\textbf{RQ3}}&\multirow{2}{*}{\textbf{RQ4}}\\
\textbf{No.}&\textbf{No.}&&&&&\\\hline
PS\addtocounter{rnum}{1}\arabic{rnum}&\cite{ABACKERLI2010}&A case study on testing software packages used in metrology&&$\checkmark$&$\checkmark$& \\\hline
PS\addtocounter{rnum}{1}\arabic{rnum}&\cite{4548407}&Software engineering tasks carried out during scientific software development&&$\checkmark$&&\\\hline
PS\addtocounter{rnum}{1}\arabic{rnum}&\cite{BagnaraCGG13}&Test case generation for floating point programs using symbolic execution&&$\checkmark$&&\\\hline
PS\addtocounter{rnum}{1}\arabic{rnum}&\cite{Carver:2007:SDE:1248820.1248886}&Case studies of scientific software development projects&&$\checkmark$&&\\\hline
PS\addtocounter{rnum}{1}\arabic{rnum}&\cite{CaverWhat2011}&Survey on computational scientists and engineers&$\checkmark$&$\checkmark$&&\\\hline
PS\addtocounter{rnum}{1}\arabic{rnum}&\cite{1045022}&Applying metamorphic testing to programs on partial differential equations &&$\checkmark$&$\checkmark$&\\\hline
PS\addtocounter{rnum}{1}\arabic{rnum}&\cite{journals/bmcbi/ChenHLX09}&Case studies on applying metamorphic testing for bioinformatics programs &&$\checkmark$&$\checkmark$&\\\hline
PS\addtocounter{rnum}{1}\arabic{rnum}&\cite{5999647}&Case studies on applying test driven development for climate models&&$\checkmark$&$\checkmark$&\\\hline
PS\addtocounter{rnum}{1}\arabic{rnum}&\cite{Cox1999339}&Using reference data sets for testing scientific software &&&$\checkmark$&\\\hline
PS\addtocounter{rnum}{1}\arabic{rnum}&\cite{DahlgrenPerformance2007}&Effectiveness of different interface contract enforcement policies for scientific components &&&$\checkmark$&\\\hline
PS\addtocounter{rnum}{1}\arabic{rnum}&\cite{Dahlgren:2005:ISS:1145319.1145341}&Partial enforcement of assertions for scientific software components&&&$\checkmark$&\\\hline
PS\addtocounter{rnum}{1}\arabic{rnum}&\cite{Davis:1981:PNP:800175.809889}&Using pseudo-oracles for testing programs without oracles&&&$\checkmark$&\\\hline
PS\addtocounter{rnum}{1}\arabic{rnum}&\cite{Drake:2005:OSD:1093627.1093634}&A case study on developing a climate system model&&&$\checkmark$&\\\hline
PS\addtocounter{rnum}{1}\arabic{rnum}&\cite{6241365}&A tool for automating the testing of scientific simulations&&$\checkmark$&&\\\hline
PS\addtocounter{rnum}{1}\arabic{rnum}&\cite{Easterbrook:2010:CCG:1882362.1882383}&Discussion on software challenges faced in climate modeling program development &&$\checkmark$&&\\\hline
PS\addtocounter{rnum}{1}\arabic{rnum}&\cite{5337646}&Ethnographic study of climate scientists who develop software&&$\checkmark$&$\checkmark$&\\\hline
PS\addtocounter{rnum}{1}\arabic{rnum}&\cite{5235131}&A unit testing framework for MATLAB programs&&$\checkmark$&$\checkmark$&\\\hline
PS\addtocounter{rnum}{1}\arabic{rnum}&\cite{gmd-4-435-2011}&A framework for automated continuous verification of numerical simulations&&&$\checkmark$&\\\hline
PS\addtocounter{rnum}{1}\arabic{rnum}&\cite{Hannay:2009:SDU:1556904.1556928}&Results of a survey conducted to identify how scientists develop and use software in their research&&$\checkmark$&&\\\hline
PS\addtocounter{rnum}{1}\arabic{rnum}&\cite{609829}&Experiments to analyze the accuracy of scientific software through static analysis and comparisons with independent implementations of the same algorithm&&$\checkmark$&$\checkmark$&\\\hline
PS\addtocounter{rnum}{1}\arabic{rnum}&\cite{328993}&N-version programming experiment conducted on scientific software&&$\checkmark$&&\\\hline
PS\addtocounter{rnum}{1}\arabic{rnum}&\cite{4135253}&Applying software quality assurance practices in a scientific software project&&$\checkmark$&&\\\hline
PS\addtocounter{rnum}{1}\arabic{rnum}&\cite{5069157}&Software engineering practices suitable for scientific software development teams identified through a case study &&$\checkmark$&$\checkmark$&\\\hline
PS\addtocounter{rnum}{1}\arabic{rnum}&\cite{4476223}&Software development process of five large scale computational science software&&&$\checkmark$&\\\hline
PS\addtocounter{rnum}{1}\arabic{rnum}&\cite{5069163}&Evaluating the effectiveness of using a small number of carefully selected test cases for testing scientific software&&$\checkmark$&$\checkmark$&\\\hline
PS\addtocounter{rnum}{1}\arabic{rnum}&\cite{DBLP:journals/bmcbi/KaneHCMKB06}&Qualitative study of agile development approaches for creating and maintaining bio-medical software&&&$\checkmark$&\\\hline
PS\addtocounter{rnum}{1}\arabic{rnum}&\cite{Kelly201147}&Comparing the effectiveness of random test cases and designed test cases for detecting faults in scientific software&&$\checkmark$&$\checkmark$&\\\hline
PS\addtocounter{rnum}{1}\arabic{rnum}&\cite{5228715}&Useful software engineering techniques for computational scientists obtained through experience of scientists who had success&&$\checkmark$&&\\\hline
PS\addtocounter{rnum}{1}\arabic{rnum}&\cite{KellyAssesing2008}&Quality assessment practices of scientists that develop computational software&&$\checkmark$&$\checkmark$&\\\hline
PS\addtocounter{rnum}{1}\arabic{rnum}&\cite{5999781}&How software engineering research can provide solutions to challenges found by scientists developing software&$\checkmark$&$\checkmark$&&$\checkmark$\\\hline
\end{tabular}
\end{table*}  

\begin{table*}
\footnotesize
\caption{Selected Primary Studies (Part 2)}
\label{tab:primaryStudies2}
\centering
\begin{tabular}{|l|l|p{0.55\textwidth}|c|c|c|c|}
\hline
\textbf{Study}&\textbf{Ref.}&\multirow{2}{*}{\textbf{Study focus}}&\multirow{2}{*}{\textbf{RQ1}}&\multirow{2}{*}{\textbf{RQ2}}&\multirow{2}{*}{\textbf{RQ3}}&\multirow{2}{*}{\textbf{RQ4}}\\
\textbf{No.}&\textbf{No.}&&&&&\\\hline
PS\addtocounter{rnum}{1}\arabic{rnum}&\cite{5467013}&A case study of applying testing activities to scientific software&&$\checkmark$&$\checkmark$&\\\hline
PS\addtocounter{rnum}{1}\arabic{rnum}&\cite{chinASurvey2007}&A survey on testing tools for scientific programs written in FORTRAN&&$\checkmark$&&\\\hline
PS\addtocounter{rnum}{1}\arabic{rnum}&\cite{doi:10.1080/0952813X.2012.695443}&A case study on using a three level testing architecture for testing scientific programs &&&$\checkmark$&\\\hline
PS\addtocounter{rnum}{1}\arabic{rnum}&\cite{4032272}&Applying metamorphic testing for image processing programs&&&$\checkmark$&\\\hline
PS\addtocounter{rnum}{1}\arabic{rnum}&\cite{Mayer05ontesting}&Using statistical oracles to test image processing applications&&&$\checkmark$&\\\hline
PS\addtocounter{rnum}{1}\arabic{rnum}&\cite{meinkeALearning2010}&A learning-based method for automatic generation of test cases for numerical programs&&&$\checkmark$&\\\hline
PS\addtocounter{rnum}{1}\arabic{rnum}&\cite{Morris_2008}&Lessons learned through code reviews of scientific programs&&$\checkmark$&&\\\hline
PS\addtocounter{rnum}{1}\arabic{rnum}&\cite{5380863}&Challenges faced by software engineers developing software for scientists in the field of molecular biology&$\checkmark$&$\checkmark$&&\\\hline
PS\addtocounter{rnum}{1}\arabic{rnum}&\cite{Murphy:2007:PRT:1292414.1292425}&A framework for randomly generating large data sets for testing machine learning applications&&$\checkmark$&$\checkmark$&\\\hline
PS\addtocounter{rnum}{1}\arabic{rnum}&\cite{Murphy_anapproach}&Methods for testing machine learning algorithms&&$\checkmark$&$\checkmark$&\\\hline
PS\addtocounter{rnum}{1}\arabic{rnum}&\cite{Murphy_propertiesof}&Applying metamorphic testing for testing machine learning applications&&&$\checkmark$&\\\hline
PS\addtocounter{rnum}{1}\arabic{rnum}&\cite{Murphy:2011:ETH:1987993.1988003}&Testing health care simulation software using metamorphic testing &&$\checkmark$&$\checkmark$&\\\hline
PS\addtocounter{rnum}{1}\arabic{rnum}&\cite{Nguyen-Hoan:2010:SSS:1852786.1852802}&Survey of scientific software developers &&$\checkmark$&$\checkmark$&\\\hline
PS\addtocounter{rnum}{1}\arabic{rnum}&\cite{gmd-5-1009-2012}&Analysis of quality of climate models in terms of defect density &&$\checkmark$&&\\\hline
PS\addtocounter{rnum}{1}\arabic{rnum}&\cite{Pitt-Francis13092008}&Applying agile development process for developing computational biology software&&$\checkmark$&$\checkmark$&\\\hline
PS\addtocounter{rnum}{1}\arabic{rnum}&\cite{Post:2004:SPM}&Lessons learned from scientific software development projects&&&$\checkmark$&\\\hline
PS\addtocounter{rnum}{1}\arabic{rnum}&\cite{6086527}&Applying variability modeling for selecting test cases when testing scientific frameworks  with large variability&&$\checkmark$&$\checkmark$&\\\hline
PS\addtocounter{rnum}{1}\arabic{rnum}&\cite{5439}&Medical software development and testing&&&$\checkmark$&\\\hline
PS\addtocounter{rnum}{1}\arabic{rnum}&\cite{4548404}&A survey to identify the characteristics of scientific software development&&$\checkmark$&$\checkmark$&\\\hline
PS\addtocounter{rnum}{1}\arabic{rnum}&\cite{sanders08challenge}&Challenges faced when testing scientific software identified through interviews carried out with scientists who develop/use scientific software&$\checkmark$&$\checkmark$&$\checkmark$&\\\hline
PS\addtocounter{rnum}{1}\arabic{rnum}&\cite{oro17671}&Study of problems arising when scientists and software engineers work together to develop scientific software&&$\checkmark$&&\\\hline
PS\addtocounter{rnum}{1}\arabic{rnum}&\cite{oro17674}&Problems in scientific software development identified through case studies in different fields&&$\checkmark$&&\\\hline
PS\addtocounter{rnum}{1}\arabic{rnum}&\cite{5069156}&Challenges faced by software engineers who develop software for scientists&&$\checkmark$&&\\\hline
PS\addtocounter{rnum}{1}\arabic{rnum}&\cite{oro18494}&Case studies on professional end-user development culture&&$\checkmark$&&\\\hline
PS\addtocounter{rnum}{1}\arabic{rnum}&\cite{SegalModels2008}&A model of scientific software development identified through multiple field studies of scientific software development&&$\checkmark$&&\\\hline
PS\addtocounter{rnum}{1}\arabic{rnum}&\cite{segalWhenSoftware2005}&A case study on applying traditional document-led development methodology for developing scientific software&&$\checkmark$&&\\\hline
PS\addtocounter{rnum}{1}\arabic{rnum}&\cite{Sletholt:2011:LRA:1985782.1985784}&Literature review and case studies on how scientific software development matches agile practices and the effects of using agile practices in scientific software development&&$\checkmark$&$\checkmark$&\\\hline
PS\addtocounter{rnum}{1}\arabic{rnum}&\cite{smithATest2007}&A test harness for numerical programs&&&$\checkmark$&\\\hline
PS\addtocounter{rnum}{1}\arabic{rnum}&\cite{2004SPIE.5423..288S}&A testing framework for conducting regression testing of scientific software&&$\checkmark$&$\checkmark$&\\\hline
PS\addtocounter{rnum}{1}\arabic{rnum}&\cite{vilkomirModeling2008}&A method for test case generation of scientific software when there are dependencies between input parameters&&$\checkmark$&$\checkmark$&\\\hline
PS\addtocounter{rnum}{1}\arabic{rnum}&\cite{Weyuker01111982}&Testing non-testable programs&&$\checkmark$&$\checkmark$&\\\hline
PS\addtocounter{rnum}{1}\arabic{rnum}&\cite{1196317}&Culture clash when applying extreme programming to develop scientific software&&$\checkmark$&$\checkmark$&\\\hline
\end{tabular}
\end{table*}

\begin{table*}
\footnotesize
\caption{Publication venues of primary studies}
\label{tab:pubVenues}
\centering
\begin{tabular}[!h]{|p{0.7\textwidth}|l|l|l|}
\hline
\textbf{Publication venue}&\textbf{Type}&\textbf{Count}&\textbf{\%}\\\hline
International Workshop on Software Engineering for Computational Science and Engineering&Workshop&7&11.3\\\hline
Computing in Science \& Engineering&Journal&7&11.3\\\hline
IEEE Software&Journal&5&8.1\\\hline
BMC Bioinformatics&Journal&2&3.2\\\hline
Geoscientific Model Development&Journal&2&3.2\\\hline
International Conference on Software Engineering and Knowledge Engineering&Conference&2&3.2\\\hline
International Journal of High Performance Computing Applications&Journal&2&3.2\\\hline
Lecture Notes in Computer Science&Book chapter&2&3.2\\\hline
Journal of the Brazilian Society of Mechanical Sciences and Engineering&Journal&1&1.6\\\hline
International Conference on Software Testing, Verification and Validation&Conference&1&1.6\\\hline
International Conference on Software Engineering&Conference&1&1.6\\\hline
Sandia National Laboratories-Technical report&Tech. report&1&1.6\\\hline
Computer Software and Applications Conference&Conference&1&1.6\\\hline
Analytica Chimica Acta&Journal&1&1.6\\\hline
International Workshop on Software Engineering for High Performance Computing System Applications&Workshop&1&1.6\\\hline
ACM '81 Conference&Conference&1&1.6\\\hline
FSE/SDP Workshop on Future of Software Engineering Research&Workshop&1&1.6\\\hline
IEEE Computational Science \& Engineering&Journal&1&1.6\\\hline
IEEE Transactions on Software Engineering&Journal&1&1.6\\\hline
EUROMICRO International Conference on Parallel, Distributed and Network-Based Processing&Conference&1&1.6\\\hline
IEEE Computer&Journal&1&1.6\\\hline
Journal of Computational Science&Journal&1&1.6\\\hline
Rutherford Appleton Laboratory-Technical report&Tech. report&1&1.6\\\hline
Journal of Experimental \& Theoretical Artificial Intelligence&Journal&1&1.6\\\hline
International Conference on Quality Software&Conference&1&1.6\\\hline
Lecture Notes in Informatics&Book chapter&1&1.6\\\hline
International Conference on e-Science&Conference&1&1.6\\\hline
International Workshop on Random testing&Conference&1&1.6\\\hline
Workshop on Software Engineering in Health Care&Workshop&1&1.6\\\hline
International Symposium on Empirical Software Engineering and Measurement&Conference&1&1.6\\\hline
Philosophical Transactions of the Royal Society A: Mathematical, Physical and Engineering Sciences&Journal&1&1.6\\\hline
Symposium on the Engineering of Computer-Based Medical Systems&Conference&1&1.6\\\hline
Conference for the Association for Software Testing&Conference&1&1.6\\\hline
Annual Meeting of the Psychology of Programming Interest Group&Conference&1&1.6\\\hline
Symposium on Visual Languages and Human-Centric Computing&Conference&1&1.6\\\hline
Computer Supported Cooperative Work&Journal&1&1.6\\\hline
Empirical Software Engineering&Journal&1&1.6\\\hline
Grid-Based Problem Solving Environments&Journal&1&1.6\\\hline
Society of Photo-Optical Instrumentation Engineers (SPIE) Conference Series&Conference&1&1.6\\\hline
International Conference on Computational Science&Conference&1&1.6\\\hline
The Computer Journal&Journal&1&1.6\\\hline
\end{tabular}
\end{table*}

\subsubsection{Data extraction and quality assessment}
We used the data extraction form in Table~\ref{tab:dataForm} to extract data from the primary studies. Many primary studies did not answer all of the questions in the data extraction form. We extracted the important information provided by the primary studies using the data extraction form. Then, depending on the type of the study, we applied the quality assessment questions in Table~\ref{tab:qualityForQuant} or Table~\ref{tab:qualityForQual} to each primary study.    

We provided `yes' and `no' answers to our quality assessment questions. We used a binary scale since we were not interested in providing a quality score for the studies~\cite{4343750}. Table~\ref{tab:QAQuan} shows the results of the quality assessment for quantitative primary studies. All the quantitative primary studies answered `yes' to the quality assessment question G1 (Are the study aims clearly stated?). Most of the quantitative primary studies answered `yes' to the quality assessment questions G2 (Are the data collection methods adequately described) and G5 (Can the study be replicated?). Table~\ref{tab:QAQuali} shows the results of the quality assessment for qualitative primary studies. All of the qualitative primary studies answered `yes' to the quality assessment question A (Are the study aims clearly stated?) and B (Does the evaluation address its stated aims and purpose?). Most of the qualitative primary studies answered `yes' to the quality assessment question D (Is enough evidence provided to support the claims?).
\begin{table*}
\footnotesize
\caption{Quality assessment results of quantitative studies}
\label{tab:QAQuan}
\centering
\begin{tabular}[!h]{|l|l|l|l|l|l|l|l|l|l|l|l|l|}
\hline
\textbf{Ref. No.}&\textbf{G1}&\textbf{S1}&\textbf{S2}&\textbf{E1}&\textbf{E2}&\textbf{G2}&\textbf{G3}&\textbf{S3}&\textbf{C1}&\textbf{E3}&\textbf{G4}&\textbf{G5}\\\hline
\cite{Drake:2005:OSD:1093627.1093634}&yes&N/A&N/A&N/A&N/A&no&no&N/A&yes&N/A&no&no\\\hline
\cite{Murphy:2011:ETH:1987993.1988003}&yes&N/A&N/A&no&no&yes&no&N/A&N/A&yes&yes&yes\\\hline
\cite{Murphy:2007:PRT:1292414.1292425}&yes&N/A&N/A&no&no&yes&no&N/A&N/A&yes&yes&yes\\\hline
\cite{Kelly201147}&yes&N/A&N/A&yes&no&yes&yes&N/A&N/A&yes&no&yes\\\hline
\cite{gmd-5-1009-2012}&yes&N/A&N/A&N/A&N/A&yes&no&N/A&yes&N/A&yes&yes\\\hline
\cite{ABACKERLI2010}&yes&N/A&N/A&N/A&N/A&yes&no&N/A&yes&N/A&no&yes\\\hline
\cite{Cox1999339}&yes&N/A&N/A&no&no&yes&yes&N/A&N/A&yes&no&yes\\\hline
\cite{Dahlgren:2005:ISS:1145319.1145341}&yes&N/A&N/A&yes&no&yes&no&N/A&N/A&yes&no&yes\\\hline
\cite{CaverWhat2011}&yes&yes&no&N/A&N/A&yes&yes&yes&N/A&N/A&no&yes\\\hline
\cite{Nguyen-Hoan:2010:SSS:1852786.1852802}&yes&yes&no&N/A&N/A&yes&no&yes&N/A&N/A&yes&yes\\\hline
\cite{Pitt-Francis13092008}&yes&N/A&N/A&N/A&N/A&yes&no&N/A&yes&N/A&no&yes\\\hline
\cite{6086527}&yes&N/A&N/A&N/A&N/A&yes&no&N/A&yes&N/A&no&yes\\\hline
\cite{5069163}&yes&N/A&N/A&yes&no&yes&no&N/A&N/A&yes&no&yes\\\hline
\cite{DahlgrenPerformance2007}&yes&N/A&N/A&yes&no&yes&no&N/A&N/A&yes&no&yes\\\hline
\cite{4032272}&yes&N/A&N/A&no&no&yes&no&N/A&N/A&yes&no&yes\\\hline
\cite{journals/bmcbi/ChenHLX09}&yes&N/A&N/A&N/A&N/A&yes&no&N/A&yes&N/A&yes&yes\\\hline
\cite{1045022}&yes&N/A&N/A&N/A&N/A&yes&no&N/A&yes&N/A&no&yes\\\hline
\cite{1196317}&yes&N/A&N/A&N/A&N/A&yes&no&N/A&yes&N/A&yes&no\\\hline
\cite{BagnaraCGG13}&yes&N/A&N/A&yes&no&yes&no&N/A&N/A&yes&no&yes\\\hline
\cite{Hannay:2009:SDU:1556904.1556928}&yes&yes&no&N/A&N/A&yes&no&yes&N/A&N/A&yes&yes\\\hline
\cite{609829}&yes&N/A&N/A&no&no&yes&yes&N/A&N/A&yes&no&no\\\hline
\cite{328993}&yes&N/A&N/A&no&no&yes&yes&N/A&N/A&yes&no&no\\\hline
\cite{meinkeALearning2010}&yes&N/A&N/A&yes&no&yes&no&N/A&N/A&yes&no&yes\\\hline
\cite{1196317}&yes&N/A&N/A&N/A&N/A&yes&no&N/A&yes&N/A&no&yes\\\hline
\multicolumn{13}{|l|}{G1: Are the study aims clearly stated?}\\
\multicolumn{13}{|l|}{S1: Was the method for collecting the sample data specified?}\\
\multicolumn{13}{|l|}{S2, E1: Is there a control group?}\\
\multicolumn{13}{|l|}{E2: Were the treatments randomly allocated?}\\
\multicolumn{13}{|l|}{G2: Are the data collection methods adequately described?}\\
\multicolumn{13}{|l|}{G3: Was there any statistical assessment of results?}\\
\multicolumn{13}{|l|}{S3: Do the observations support the claims?}\\
\multicolumn{13}{|l|}{C1, E3: Is there enough evidence provided to support the claims?}\\
\multicolumn{13}{|l|}{G4: Are threats to validity and/or limitations reported?}\\
\multicolumn{13}{|l|}{G5: Can the study be replicated?}\\\hline
\end{tabular}
\end{table*} 

\begin{table*}
\footnotesize
\caption{Quality assessment results of qualitative studies}
\label{tab:QAQuali}
\centering
\begin{tabular}[!h]{|l|p{0.4in}|p{0.4in}|p{0.4in}|p{0.4in}|p{0.4in}|}
\hline
\textbf{Ref. No.}&\textbf{A}&\textbf{B}&\textbf{C}&\textbf{D}&\textbf{E}\\\hline
\cite{Post:2004:SPM}&yes&yes&yes&yes&yes\\\hline
\cite{5439}&yes&yes&no&yes&no\\\hline
\cite{segalWhenSoftware2005}&yes&yes&no&yes&no\\\hline
\cite{5380863}&yes&yes&yes&yes&no\\\hline
\cite{4548407}&yes&yes&yes&yes&no\\\hline
\cite{5467013}&yes&yes&no&yes&no\\\hline
\cite{5999781}&yes&yes&yes&yes&no\\\hline
\cite{2004SPIE.5423..288S}&yes&yes&no&yes&no\\\hline
\cite{4135253}&yes&yes&yes&yes&no\\\hline
\cite{chinASurvey2007}&yes&yes&no&yes&no\\\hline
\cite{Easterbrook:2010:CCG:1882362.1882383}&yes&yes&no&yes&no\\\hline
\cite{KellyAssesing2008}&yes&yes&yes&yes&no\\\hline
\cite{4548404}&yes&yes&yes&yes&no\\\hline
\cite{5999647}&yes&yes&no&yes&no\\\hline
\cite{6241365}&yes&yes&yes&yes&yes\\\hline
\cite{smithATest2007}&yes&yes&yes&yes&no\\\hline
\cite{doi:10.1080/0952813X.2012.695443}&yes&yes&yes&yes&no\\\hline
\cite{Murphy_propertiesof}&yes&yes&yes&yes&yes\\\hline
\cite{4476223}&yes&yes&yes&yes&yes\\\hline
\cite{Mayer05ontesting}&yes&yes&yes&yes&yes\\\hline
\cite{Davis:1981:PNP:800175.809889}&yes&yes&no&yes&no\\\hline
\cite{oro17671}&yes&yes&yes&yes&yes\\\hline
\cite{oro18494}&yes&yes&yes&yes&yes\\\hline
\cite{Carver:2007:SDE:1248820.1248886}&yes&yes&yes&yes&no\\\hline
\cite{oro17674}&yes&yes&yes&yes&no\\\hline
\cite{Murphy_anapproach}&yes&yes&yes&yes&yes\\\hline
\cite{5069157}&yes&yes&yes&yes&no\\\hline
\cite{DBLP:journals/bmcbi/KaneHCMKB06}&yes&yes&yes&yes&no\\\hline
\cite{gmd-4-435-2011}&yes&yes&yes&yes&no\\\hline
\cite{5228715}&yes&yes&no&yes&no\\\hline
\cite{Morris_2008}&yes&yes&yes&yes&yes\\\hline
\cite{5337646}&yes&yes&yes&yes&yes\\\hline
\cite{5235131}&yes&yes&no&no&no\\\hline
\cite{sanders08challenge}&yes&yes&yes&yes&yes\\\hline
\cite{5069156}&yes&yes&yes&yes&no\\\hline
\cite{SegalModels2008}&yes&yes&yes&yes&no\\\hline
\cite{Sletholt:2011:LRA:1985782.1985784}&yes&yes&yes&yes&yes\\\hline
\cite{Weyuker01111982}&yes&yes&no&no&no\\\hline
\multicolumn{6}{|l|}{A: Are the study aims clearly stated?}\\
\multicolumn{6}{|l|}{B: Does the evaluation address its stated aims and purpose?}\\
\multicolumn{6}{|l|}{C: Is sample design/target selection of cases/documents defined?}\\
\multicolumn{6}{|l|}{D: Is enough evidence provided to support the claims?}\\
\multicolumn{6}{|l|}{E: Can the study be replicated?}\\\hline
\end{tabular}
\end{table*}

\subsection{Reporting the review}

Data extracted from the 62 primary papers were used to formulate answers to the four research questions given in Section~\ref{sec:RQs}. We closely followed guidelines provided by Kitchenham~\cite{citeulike:1191541} when preparing the SLR report.

\section{Results}
\label{sec:reviewReport}
We use the selected primary papers to provide answers to the research questions.
\subsection{RQ1: How is scientific software defined in the literature?}


\emph{Scientific software} is defined in various ways. Sanders \textit{et al.}~\cite{sanders08challenge} use the definition provided by Kreyman \textit{et al.}~\cite{Kreyman_inspectionprocedures}: ``Scientific software is software with a large computational component and provides data for decision support.'' Kelly \textit{et al.} identified two types of scientific software~\cite{5999781}:
\begin{description}
 \item[(1)] End user application software that is written to achieve scientific objectives (e.g., Climate models).
 \item[(2)] Tools that support writing code that express a scientific model and the execution of scientific code (e.g., Automated software testing tool for MATLAB~\cite{5235131}).
\end{description}
An orthogonal classification is given by Carver \textit{et al.}~\cite{CaverWhat2011}: 
\begin{description}
	\item[(1)] Research software written with the goal of publishing papers.
	\item[(2)] Production software written for real users (e.g. Climate models).
\end{description}

Scientific software is developed by scientists themselves or by multi-disciplinary teams, where a team consists of scientists and professional software developers. A scientist will generally be the person in charge of a scientific software development project~\cite{5380863}.  

We encountered software that helps to solve a variety of scientific problems. We present the details of software functionality, size and the programing languages in Table~\ref{tab:sw}. None of the primary studies reported the complexity of the software in terms of measurable unit such as coupling, cohesion, or cyclomatic complexity.

\begin{table*}[!ht]
  \centering
  \footnotesize
   \caption{Details of scientific software listed in primary studies}
  \begin{tabular}{|p{0.065\textwidth}|p{0.65\textwidth}|p{0.13\textwidth}|p{0.1\textwidth}|}
  \hline
    \textbf{Ref. No.}&\textbf{Description}&\textbf{Programing language}&\textbf{Size}\\\hline
    \cite{5439}&Medical software (e.g. software for blood chemistry analyzer and medical image processing system)&N/S&N/S\\\hline
    \cite{Post:2004:SPM}&Nuclear weapons simulation software&FORTRAN&500 KLOC\\\hline
    \cite{Drake:2005:OSD:1093627.1093634,Easterbrook:2010:CCG:1882362.1882383}&Climate modeling  software&N/S&N/S\\\hline
    \cite{segalWhenSoftware2005}&Embedded software for spacecrafts&N/S&N/S\\\hline
    \cite{5380863}&Software developed for space scientists and biologists
&N/S&N/S\\\hline
    \cite{4548407}&Control and data acquisition software for Synchrotron Radiation Source (SRS) experiment stations&Java&N/S\\\hline
    \cite{Murphy:2011:ETH:1987993.1988003}&Health care simulation software(e.g. discreet event simulation engine and insulin titration algorithm simulation)&Java, MATLAB&N/S\\\hline
    \cite{Murphy:2007:PRT:1292414.1292425}&Machine learning ranking algorithm implementations&Perl, C&N/S\\\hline
    \cite{gmd-5-1009-2012}&Climate modeling software&FORTRAN, C&400 KLOC\\\hline
    \cite{5467013}&Astronomy software package&MATLAB, C++&10 KLOC\\\hline
    \cite{ABACKERLI2010}&Software packages providing uncertainty estimates for tri-dimensional measurements&N/S&N/S\\\hline
    \cite{2004SPIE.5423..288S}&Implementation of a time dependent simulation of a complex physical system&N/S&N/S\\\hline
    \cite{Dahlgren:2005:ISS:1145319.1145341}&Implementation of scientific mesh traversal algorithms&N/S&38-50 LOC\\\hline
    \cite{4135253}&Implementations of parallel solver algorithms and libraries for large scale, complex, multi physics engineering and scientific applications&N/S&N/S\\\hline
    \cite{Pitt-Francis13092008}&Software for cardiac modeling in computational biology&C++, Python&50 KLOC\\\hline
    \cite{gmd-4-435-2011}&Numerical simulations in geophysical fluid dynamics&N/S&N/S\\\hline 
    \cite{6086527}&Program for solving partial differential equations&C++&250 KLOC\\\hline
    \cite{5999647}&Calculates the trajectory of the billions of air particles in the atmosphere&C++&N/S\\\hline
    \cite{5999647}&Implementation of a numerical model that simulates the growth of virtual snow flakes&C++&N/S\\\hline
    \cite{DahlgrenPerformance2007}&Implementations of mesh traversal algorithms&N/S&N/S\\\hline
    \cite{4032272}&Image processing application&N/S&N/S\\\hline
    \cite{journals/bmcbi/ChenHLX09}&Bioinformatics program for analyzing and simulating gene regulatory networks and mapping short sequence reads to a reference genome&N/S&N/S\\\hline
    \cite{Murphy:2007:PRT:1292414.1292425,Murphy_anapproach}&Implementations of machine learning algorithms&N/S&N/S\\\hline
    \cite{4476223}&Simulations in solid mechanics, fluid mechanics and combustion&C, C++, FORTRAN&100-500 KLOC\\\hline
    \cite{1196317}&Program to evaluate the performance of a numerical scheme to solve a model advection-diffusion problem&Ruby&2.5 KLOC\\\hline
    \cite{Mayer05ontesting}&Implementation of dilation of binary images&N/S&N/S\\\hline
    \cite{oro18494}&Infrastructure software for the structural protein community&N/S&N/S\\\hline
    \cite{Carver:2007:SDE:1248820.1248886}&Performance prediction software for a product that otherwise requires large, expensive and potentially dangerous empirical tests for performance evaluation&FORTRAN, C&405 KLOC\\\hline
    \cite{Carver:2007:SDE:1248820.1248886}&Provide computational predictions to analyze the manufacturing process of composite material products&C++, C&134 KLOC\\\hline
    \cite{Carver:2007:SDE:1248820.1248886}&Simulation of material behavior when placed under extreme stress&FORTRAN&200 KLOC\\\hline
    \cite{Carver:2007:SDE:1248820.1248886}&Provide real-time processing of sensor data&C++, MATLAB&100 KLOC\\\hline
    \cite{Carver:2007:SDE:1248820.1248886}&Calculate the properties of molecules using computational quantum mechanical models&FORTRAN&750 KLOC\\\hline
    \cite{BagnaraCGG13}&Program for avoiding collisions in unmanned aircrafts&C&N/S\\\hline
    \cite{5069157}&Numerical libraries to be used by computational science and engineering software projects&N/S&N/S\\\hline
  \end{tabular}
  \label{tab:sw}
\end{table*}
\subsection{RQ2: Are there special characteristics or faults in scientific software or its development that make testing difficult?}

\label{sec:challenges}
We found characteristics that fall into two main categories 1) Testing challenges that occur due to characteristics of scientific software, and 2) Testing challenges that occur due to cultural differences between scientists and the software engineering community. Below we describe these challenges:
\begin{enumerate}
	\item Testing challenges that occur due to characteristics of scientific software: These challenges can be further categorized according to the specific testing activities where they pose problems.
		\begin{enumerate}
		\item Challenges concerning test case development:
			\begin{enumerate}
			\item Identifying critical input domain boundaries \textit{a priori} is difficult due to the complexity of the software, round-off error effects, and complex computational behavior. This makes it difficult to apply techniques such as equivalence partitioning to reduce the number of test cases~\cite{4548404,Kelly201147,Carver:2007:SDE:1248820.1248886}.
			\item Manually selecting a sufficient set of test cases is challenging due to the large number of input parameters and values accepted by some scientific software~\cite{vilkomirModeling2008}.
			\item When testing scientific frameworks at the system level, it is difficult to choose a suitable set of test cases from the large number of available possibilities ~\cite{6086527}.
			\item Some scientific software lacks real world data that can be used for testing~\cite{Murphy:2007:PRT:1292414.1292425}.
			\item Execution of some paths in scientific software are dependent on results of floating point calculations. Finding test data to execute such program paths is challenging~\cite{BagnaraCGG13}.
			 \item Some program units (functions, subroutines, methods) in scientific software contain so many decisions that testing is impractical~\cite{Morris_2008}.
			 \item Difficulties in replicating the physical context where the scientific code is suppose to work can make comprehensive testing impossible~\cite{segalWhenSoftware2005}.
			\end{enumerate}
		\item Challenges towards producing expected test case output values (Oracle problems): Software testing requires an oracle, a mechanism for checking whether the program under test produces the expected output when executed using a set of test cases. Obtaining reliable oracles for scientific programs is challenging~\cite{sanders08challenge}. Due to the lack of suitable oracles it is difficult to detect subtle faults in scientific code~\cite{5228715}. The following characteristics of scientific software make it challenging to create a test oracle:   
			 \begin{enumerate}
			  \item Some scientific software is written to find answers that are previously unknown. Therefore only approximate solutions might be available~\cite{Easterbrook:2010:CCG:1882362.1882383,Murphy:2011:ETH:1987993.1988003,Weyuker01111982,Carver:2007:SDE:1248820.1248886,5999781}. 
			  \item It is difficult to determine the correct output for software written to test scientific theory that involves complex calculations or simulations. Further, some programs produce complex outputs making it difficult to determine the expected output~\cite{Sletholt:2011:LRA:1985782.1985784,sanders08challenge,Murphy_anapproach,journals/bmcbi/ChenHLX09,KellyAssesing2008,Pitt-Francis13092008,Hannay:2009:SDU:1556904.1556928,Weyuker01111982,oro17671}. 
			  \item Due to the inherent uncertainties in models, some scientific programs do not give a single correct answer for a given set of inputs. This makes determining the expected behavior of the software a difficult task, which may depend on a domain expert's opinion~\cite{ABACKERLI2010}.
			  \item Requirements are unclear or uncertain up-front due to the exploratory nature of the software. Therefore developing oracles based on requirements is not commonly done~\cite{Sletholt:2011:LRA:1985782.1985784,Nguyen-Hoan:2010:SSS:1852786.1852802,Hannay:2009:SDU:1556904.1556928,4135253}.
			  \item Choosing suitable tolerances for an oracle when testing numerical programs is difficult due to the involvement of complex floating point computations~\cite{Pitt-Francis13092008,Kelly201147,5467013,5999647}.
			\end{enumerate}
		\item Challenges towards test execution:
			\begin{enumerate}
			\item Due to long execution times of some scientific software, running a large number of test cases to satisfy specific coverage criteria is not feasible~\cite{Kelly201147}.
			\end{enumerate}
		\item Challenges towards test result interpretation:
		\begin{enumerate}
			\item Faults can be masked by round-off errors, truncation errors and model simplifications~\cite{Kelly201147,328993,Hannay:2009:SDU:1556904.1556928,1045022,5999647}.
			\item A limited portion of the software is regularly used. Therefore, less frequently used portions of the code may contain unacknowledged errors~\cite{gmd-5-1009-2012,chinASurvey2007}.
			\item Scientific programs contain a high percentage of duplicated code~\cite{Morris_2008}.
		\end{enumerate}
	\end{enumerate}
		
	\item Testing challenges that occur due to cultural differences between scientists and the software engineering community: Scientists generally play leading roles in developing scientific software.  
	\begin{enumerate}
		\item Challenges due to limited understanding of testing concepts:
		\begin{enumerate}
			\item Scientists view the code and the model that it implements as inseparable entities. Therefore they test the code to assess the model and not necessarily to check for faults in the code~\cite{KellyAssesing2008,chinASurvey2007,4548404,sanders08challenge}.
			\item Scientist developers focus on the scientific results rather than the quality of the software~\cite{5337646,Carver:2007:SDE:1248820.1248886}.
			\item The value of the software is underestimated~\cite{oro17671}.
			\item Definitions of verification and validation are not consistent across the computational science and engineering communities~\cite{5069163}.
			\item Developers (scientists) have little or no training in software engineering~\cite{Easterbrook:2010:CCG:1882362.1882383,5337646,Hannay:2009:SDU:1556904.1556928,Carver:2007:SDE:1248820.1248886,CaverWhat2011}.
			\item Requirements and software evaluation activities are not clearly defined for scientific software~\cite{SegalModels2008,oro18494}.
			\item Testing is done only with respect to the initial specific scientific problem addressed by the code. Therefore the reliability of results when applied to a different problem cannot be guaranteed~\cite{5380863}.
			\item Developers are unfamiliar with testing methods~\cite{5235131,Hannay:2009:SDU:1556904.1556928}.
		\end{enumerate}
		\item Challenges due to limited understanding of testing process
		\begin{enumerate}
			\item Management and budgetary support for testing may not be provided~\cite{Nguyen-Hoan:2010:SSS:1852786.1852802,4135253,oro18494}. 
			\item Since the requirements are not known up front, scientists may adopt an agile philosophy for development. However, they do not use standard agile process models~\cite{5337646}. As a result, unit testing and acceptance testing are not carried out properly.
			\item Software development is treated as a secondary activity resulting in a lack of recognition for the skills and knowledge required for software development~\cite{oro17674}.
			\item Scientific software does not usually have a set of written or agreed set of quality goals~\cite{Morris_2008}. 
			\item Often only ad-hoc or unsystematic testing methods are used~\cite{sanders08challenge,oro17674}.
			\item Developers view testing as a task that should be done late during software development~\cite{5069157}.
		\end{enumerate}
		\item Challenges due to not applying known testing methods
		\begin{enumerate}
			\item The wide use of FORTRAN in the scientific community makes it difficult to utilize many testing tools from the software engineering community~\cite{chinASurvey2007,4548404,5337646}.  
			\item Unit testing is not commonly conducted when developing scientific software~\cite{1196317,6241365}. For example, Clune \textit{et al.} find that unit testing is almost non-existent in the climate modeling community~\cite{5999647}. Reasons for the lack of unit testing include the following:
    \begin{itemize}
    	\item There are misconceptions about the difficulty and benefits of implementing unit tests among scientific software developers~\cite{5999647}. 
    	\item The legacy nature of scientific code makes implementing unit tests challenging~\cite{5999647}.
    	\item The internal code structure is hidden~\cite{2004SPIE.5423..288S}. 
    	\item The importance of unit testing is not appreciated by scientist developers~\cite{5069156}.
    \end{itemize}
     \item Scientific software developers are unaware of the need for and the method of applying verification testing~\cite{sanders08challenge}.
     \item There is a lack of automated regression and acceptance testing in some scientific programs~\cite{CaverWhat2011}.
		\end{enumerate}
    
	\end{enumerate}
\end{enumerate}


The following specific faults are reported in the selected primary studies:   
\begin{itemize}
	\item Incorrect use of a variable name~\cite{1045022}.
	\item Incorrectly reporting hardware failures as faults due to ignored exceptions~\cite{Morris_2008}.
	\item One-off errors~\cite{609829}.
\end{itemize}

\subsection{RQ3: Can we use existing testing methods (or adapt them) to test scientific software effectively?}

\begin{figure}[h!]
  \caption{Percentage of studies that applied different testing methods}
  \label{fig:testingTypePerc}
  \centering
    \includegraphics[width=0.6\textwidth]{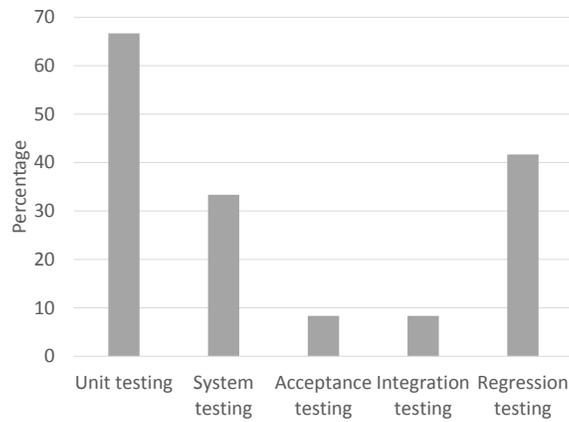}
\end{figure}

\textbf {Use of testing at different abstraction levels and for different testing purposes.} Several primary studies reported conducting testing at different abstraction levels: unit testing, integration testing and system testing. In addition some studies reported the use of acceptance testing and regression testing. Out of the 62 primary studies, 12 studies applied at least one of these testing methods. Figure~\ref{fig:testingTypePerc} shows the percentage of studies that applied each testing method out of the 12 studies. Unit testing was the most common testing method reported among the 12 studies.  

Figure~\ref{fig:testingTypeCounts} displays the percentage of the number of testing methods applied by the 12 studies. None of the studies applied four or more testing methods. Out of the 12 studies, 8 (67\%) mention applying only one testing method. Below we describe how these testing methods were applied when testing scientific software:  
\begin{figure}[h!]
  \caption{Number of testing methods applied by the studies}
  \label{fig:testingTypeCounts}
  \centering
    \includegraphics[width=0.6\textwidth]{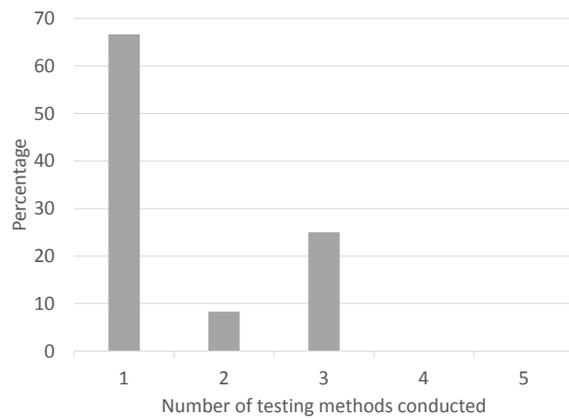}
\end{figure}   
  
\begin{enumerate}
    \item Unit testing: Several studies report that unit testing was used to test scientific programs~\cite{DBLP:journals/bmcbi/KaneHCMKB06,gmd-4-435-2011,Drake:2005:OSD:1093627.1093634,4548407,5467013,doi:10.1080/0952813X.2012.695443}. Clune \textit{et al.} describe the use of refactoring to extract testable units when conducting unit testing on legacy code~\cite{5999647}. They identified two faults using unit testing that could not be discovered by system testing. Only two studies used a unit testing framework to apply automated unit testing~\cite{DBLP:journals/bmcbi/KaneHCMKB06,4548407} and both of these studies used JUnit\footnote{http://junit.org/}. In addition, Eddins~\cite{5235131} developed a unit testing framework for MATLAB. We did not find evidence of the use of any other unit testing frameworks.    
    
    \item Integration testing: We found only one study that applied integration testing to ensure that all components work together as expected~\cite{Drake:2005:OSD:1093627.1093634}.
    
    \item System testing: Several studies report the use of system testing~\cite{DBLP:journals/bmcbi/KaneHCMKB06,gmd-4-435-2011,5439}. In particular, the climate modeling community makes heavy use of system testing~\cite{5999647}. 
    
    \item Acceptance testing: We found only one study that reports on acceptance testing conducted by the users to ensure that programmers have correctly implemented the required functionality~\cite{DBLP:journals/bmcbi/KaneHCMKB06}. One reason acceptance testing is rarely used is that the scientists who are developing the software are often also the users. 
    

    \item Regression testing: Several studies describe the use of regression testing to compare the current output to previous outputs to identify faults introduced when the code is modified~\cite{gmd-4-435-2011,Drake:2005:OSD:1093627.1093634,4476223,2004SPIE.5423..288S}. Further, Smith developed a tool for assisting regression testing~\cite{smithATest2007}. This tool allows testers to specify the variable values to be compared and tolerances for comparisons.
    
\end{enumerate}

\textbf{Techniques used to overcome oracle problems.} Previously we described several techniques used to test programs that do not have oracles~\cite{SE-CSE13p48}. In addition, several studies propose techniques to alleviate the oracle problem:
	\begin{enumerate}
    	\item A \emph{pseudo oracle} is an independently developed program that fulfills the same specification as the program under test~\cite{ABACKERLI2010,Nguyen-Hoan:2010:SSS:1852786.1852802,gmd-4-435-2011,5337646,Post:2004:SPM,sanders08challenge,Weyuker01111982,Davis:1981:PNP:800175.809889,609829}. For example, Murphy \textit{et al.} used pseudo oracles for testing a machine learning algorithm~\cite{Murphy_anapproach}.\\ 
    	\textbf{Limitations}: A pseudo oracle may not include some special features/treatments available in the program under test and it is difficult to decide whether the oracle or the program is faulty when the answers do not agree~\cite{1045022}. Pseudo oracles make the assumption that independently developed reference models will not result in the same failures. But Brilliant \textit{et al.} found that even independently developed programs might produce the same failures~\cite{44387}.

    	\item Solutions obtained analytically can serve as oracles. Using analytical solutions is sometimes preferred over pseudo oracles since they can identify common algorithmic errors among the implementations. For example, a theoretically calculated rate of convergence can be compared with the rate produced by the code to check for faults in the program~\cite{ABACKERLI2010,KellyAssesing2008,gmd-4-435-2011}. \\
    \textbf{Limitations}: Analytical solutions may not be available for every application~\cite{1045022} and may not be accurate due to human errors~\cite{sanders08challenge}.
    
    \item Experimentally obtained results can be used as oracles~\cite{ABACKERLI2010,KellyAssesing2008,Nguyen-Hoan:2010:SSS:1852786.1852802,Post:2004:SPM,sanders08challenge,doi:10.1080/0952813X.2012.695443}.\\
    \textbf{Limitations}: It is difficult to determine whether an error is due to a fault in the code or due to an error made during the model creation~\cite{1045022}. In some situations experiments cannot be conducted due to high cost, legal or safety issues~\cite{Carver:2007:SDE:1248820.1248886}. 
    
    \item Measurements values obtained from natural events can be used as oracles.\\ 
    \textbf{Limitations}: Measurements may not be accurate and are usually limited due to the high cost or danger involved in obtaining them~\cite{KellyAssesing2008,4548404}.
    
    \item Using the professional judgment of scientists~\cite{4548404,5467013,5069163,sanders08challenge}\\
    \textbf{Limitations}: Scientists can miss faults due to misinterpretations and lack of data. In addition, some faults can produce small changes in the output that might be difficult to identify~\cite{5069163}. Further, the scientist may not provide objective judgments~\cite{sanders08challenge}. 
    
    \item Using simplified data that so the correctness can be determined easily~\cite{Weyuker01111982}.\\
    \textbf{Limitations}: It is not sufficient to test using only simple data; simple test cases may not uncover faults such as round-off problems, truncation errors, overflow conditions, etc~\cite{4476223}. Further such tests do not represent how the code is actually used~\cite{sanders08challenge}.
    
    \item Statistical oracle: verifies statistical characteristics of test results~\cite{Mayer05ontesting}. \\
    \textbf{Limitations}: Decisions by a statistical oracle may not always be correct. Further a statistical oracle cannot decide whether a single test case has passed or failed~\cite{Mayer05ontesting}.
    
    \item Reference data sets: Cox \textit{et al.} created reference data sets based on the functional specification of the program that can be used for black-box testing of scientific programs~\cite{Cox1999339}.\\
    \textbf{Limitations}: When using reference data sets, it is difficult to determine whether the error is due to using unsuitable equations or due to a fault in the code.   
    
    \item \emph{Metamorphic testing (MT)} was introduced by Chen \textit{et al.}~\cite{Chen_Cheung_Yiu_1998} as a way to test programs that do not have
oracles. MT operates by checking whether a program under test behaves according to an expected set of properties known as \emph{metamorphic relations}. A metamorphic relation specifies how a particular change to the input of the program should change the output. MT was used for testing scientific applications in different areas such as machine learning applications~\cite{Xie2011544,Murphy_propertiesof}, bioinformatics programs~\cite{journals/bmcbi/ChenHLX09}, programs solving partial differential equations~\cite{1045022} and image processing applications~\cite{4032272}. When testing programs solving partial differential equations, MT uncovered faults that cannot be uncovered by special value testing~\cite{1045022}. MT can be applied to perform both unit testing and system testing. Murphy \textit{et al.} developed a supporting framework for conducting metamorphic testing at the function level~\cite{Murphy:2009:UJR:1547558.1548230}. They used the Java Modeling Language (JML) for specifying the 
metamorphic relations and automatically generating test code using the provided specifications. \emph{Statistical Metamorphic testing (SMT)} is a technique for testing non-deterministic programs that lack oracles~\cite{4385527}. Guderlei \textit{et al.} applied SMT for testing an implementation of the inverse cumulative distribution function of the normal distribution~\cite{4385527}. Further, SMT was applied for testing non-deterministic health care simulation software~\cite{Murphy:2011:ETH:1987993.1988003} and a stochastic optimization program~\cite{5463648}.\\
\textbf{Limitations}: Enumerating a set of metamorphic relations that should be satisfied by a program is a critical initial task in applying metamorphic testing. A tester or developer has to manually identify metamorphic relations using her knowledge of the program under test; this manual process can miss some important metamorphic relations that could reveal faults. Recently we proposed a novel technique based on machine learning for automatically detecting metamorphic relations~\cite{KanewalaBiemanISSTA13}.
    \end{enumerate}

As noted in Section~\ref{sec:challenges}, selecting suitable tolerances for oracles is another challenge. Kelly \textit{et al.} experimentally found that reducing the tolerance in an oracle increases the ability to detect faults in the code~\cite{Kelly201147}. Clune \textit{et al.} found that breaking the algorithm into small steps and testing the steps independently reduced the compounding effects of truncation and round-off errors~\cite{5999647}. \\     
    
\textbf{Test case creation and selection.} Several methods can help to overcome the challenges in test case creation and selection:
\begin{enumerate}
\item Hook \textit{et al.} found that many faults can be identified by a small number of test cases that push the boundaries of the computation represented by the code~\cite{5069163}. Following this, Kelly \textit{et al.} found that random tests combined with specially designed test cases to cover the parts of code uncovered by the random tests are effective in identifying faults~\cite{Kelly201147}. Both of these studies used MATLAB functions in their experiments. 

\item Randomly generated test cases were used with metamorphic testing to automate the testing of image processing applications~\cite{4032272}.

\item Vilkomir \textit{et al.} developed a method for automatically generating test cases when a scientific program has many input parameters with dependencies~\cite{vilkomirModeling2008}. Vilkomir \textit{et al.} represent the input space as a directed graph. Input parameters are represented by the nodes in the graph. Specific values of the parameters and the probability of a parameter taking that value are represented by arcs. Dependencies among input parameter values are handled by splitting/merging nodes. This method  creates a model which satisfies the probability law of Markov chains. Valid test cases can be automatically generated by taking a path in this directed graph. This model also provides the ability to generate random and weighted test cases according to the likelihood of taking the parameter values.   

\item Bagnara \textit{et al.} used symbolic execution to generate test data for floating point programs~\cite{BagnaraCGG13}. This method generates test data to traverse program paths that involve floating point computations. 

\item Meinke \textit{et al.} developed a technique for Automatic test case generation for numerical software based on learning based testing (LBT)~\cite{meinkeALearning2010}. The authors first created a polynomial model as an abstraction of the program under test. Then the test cases are generated by applying a satisfiability algorithm to the learned model.

\item Parameterized random data generation is a technique described by Murphy \textit{et al.}~\cite{Murphy:2007:PRT:1292414.1292425} for creating test data for machine learning applications. This method randomly generates data sets using properties of equivalence classes.

\item Remmel \textit{et al.} developed a regression testing framework for a complex scientific framework~\cite{6086527}. They took a software product line engineering (SPLE) approach to handle the large variability of the scientific framework. They developed a variability model to represent this variability and used the model to derive test cases while making sure necessary variant combinations are covered. This approach requires that scientists help to identify infeasible combinations. 
\end{enumerate}   

\textbf{Test coverage information.} Only two primary studies mention the use of some type of test coverage information~\cite{DBLP:journals/bmcbi/KaneHCMKB06,4548407}. Kane \textit{et al.} found that while some developers were interested in measuring statement coverage, most of the developers were interested in covering the significant functionality of the program~\cite{DBLP:journals/bmcbi/KaneHCMKB06}. Ackroyd \textit{et al.}~\cite{4548407} used the Emma tool to measure test coverage. 

\textbf{Assertion checking.} Assertion checking can be used to ensure the correctness of plug-and-play scientific components. But assertion checking introduces a performance overhead. Dahlgren \textit{et al.} developed an assertion selection system to reduce performance overhead for scientific software~\cite{Dahlgren:2005:ISS:1145319.1145341,DahlgrenPerformance2007}.

\textbf{Software development process.} Several studies reported that using agile practices for developing scientific software improved testing activities~\cite{Sletholt:2011:LRA:1985782.1985784,Pitt-Francis13092008,1196317}. Some projects have used test-driven development (TDD), where test are written to check the functionality before the code is written. But adopting this approach could be a cultural challenge since primary studies report that TDD can delay the initial development of functional code~\cite{5069157,4548407}.


\subsection{RQ4: Are there any challenges that could not be answered by existing techniques?}

Only one primary paper directly provided answers to RQ4. Kelly \textit{et al.}~\cite{5999781} describes oracle problems as key problems to solve and the need for research on performing effective testing without oracles. We did not find other answers to this research question.


\section{Discussion}
\label{sec:discussion}
\subsection{Principal findings}

The goal of this systematic literature review is to identify specific challenges faced when testing scientific software, how the challenges have been met, and any unsolved challenges. The principal findings of this review are the following:

\begin{enumerate}
	\item The main challenges in testing scientific software can be grouped into two main categories. 
		\begin{itemize}
		\item Testing challenges that occur due to characteristics of scientific software.
			\begin{itemize}
			\item Challenges concerning test case development such as a lack of real world data and difficulties in replicating the physical context where the scientific code is suppose to work.
			\item Oracle problems mainly arise because scientific programs are either written to find answers that are previously unknown or they perform complex calculations so that it is difficult to determine the correct output. 30\% of the primary studies reported the oracle problems as challenges for conducting testing.
			\item Challenges towards test execution such as difficulties in running test cases to satisfy a coverage criteria due to long execution times.
			\item Challenges towards test result interpretation such as round-off errors, truncation errors and model simplifications masking faults in the code. 
			\end{itemize}
		\item Testing challenges that occur due to cultural differences between scientists and the software engineering community.
		\begin{itemize}
		\item Challenges due to limited understanding of testing concepts such as viewing the code and the model that it implements as inseparable entities.
		\item Challenges due to limited understanding of testing processes resulting in the use of ad-hoc or unsystematic testing methods.
		\item Challenges due to not applying known testing methods such as unit testing.
		\end{itemize}
		\end{itemize}
	\item We discovered how certain techniques can be used to overcome some of the testing challenges posed by scientific software development.
		\begin{itemize}
		\item Pseudo oracles, analytical solutions, experimental results, measurement values, simplified data and professional judgment are widely used as solutions to oracle problems in scientific software. But we found no empirical studies evaluating the effectiveness of these techniques in detecting subtle faults. New techniques such as metamorphic testing have been applied and evaluated for testing scientific software in research studies. But we found no evidence that such techniques are actually used in practice.
		\item Traditional techniques such as random test case generation were applied to test scientific software after applying modifications to consider equivalence classes. In addition, studies report the use of specific techniques to perform automatic test case generation for floating point programs. These techniques were only applied to a narrow set of programs. The applicability of these techniques in practice needs to be investigated.
		\item When considering unit, system, integration, acceptance and regression testing, very few studies applied more than one type of testing to their programs. We found no studies that applied more than three of these testing techniques.
		\item Only two primary studies evaluated some type of test coverage information during the testing process.            
		\end{itemize}
	\item Research from the software engineering community can help to improve the testing process, by investigating how to perform effective testing for programs with oracle problems.

\end{enumerate}
\subsection{Techniques potentially useful in scientific software testing}

Oracle problems are key problems to solve. Research on performing effective testing without oracles is needed~\cite{5999781}. Techniques such as property based testing and data redundancy can be used when an oracle is not available~\cite{Ammann:2008:IST:1355340}. Assertions can be used to perform property based testing within the source code~\cite{SE-CSE13p48}. Another potential approach is to use a \emph{golden run}~\cite{6258305}. With a golden run, an execution trace is generated during a failure free execution of an input. Then this execution trace is compared with execution traces obtained when executing the program with the same input when a failure is observed. By comparing the golden run and the faulty execution traces the robustness of the program is determined. One may also apply model based testing, but model based testing requires well-defined and stable requirements to develop the model. But with most scientific software, requirements are constantly changing, which can make it difficult to apply model based testing. We did not find applications of property based testing, data redundancy, golden run, and model based testing to test scientific software in the primary studies. In addition, research on test case selection and test data adequacy has not considered the effect of the oracle used. Often perfect oracles are not available for scientific programs. Therefore developing test selection/creation techniques that consider the characteristics of the oracle used for testing will be useful. 

Metamorphic testing is a promising testing technique to address the oracle problem. Metamorphic testing can be used to perform both unit and system testing. But identifying metamorphic relations that should be satisfied by a program is challenging. Therefore techniques that can identify metamorphic relations for a program are needed~\cite{KanewalaBiemanISSTA13}.   

Only a few studies applied new techniques developed by the software engineering community to overcome some of the common testing challenges. For example none of the primary studies employ test selection techniques to select test cases, even though running a large number of test cases is difficult due to the long execution times of scientific software. But many test selection techniques assume a perfect oracle, and thus will not work well for most scientific programs.  

Several studies report that scientific software developers used regression testing during the development process. But we could not determine if regression testing was automated or whether any test case prioritizing techniques were used. In addition we only found two studies that used unit testing frameworks to conduct unit testing. Both of these studies report the use of the JUnit framework for Java programs. None of the primary studies report information regarding how unit testing was conducted for programs written in other languages.

One of the challenges of testing scientific programs is duplicated code. Even though a fault is fixed in a single location, the same fault may exist in other locations and those faults can go undetected when duplicated code is present. Automatic clone detection techniques would be useful to  find duplicated code especially when dealing with legacy code. 

\subsection{Strengths and weaknesses of the SLR}
Primary studies that provided the relevant information for this literature review were identified thorough a key word based search on three databases. The search found relevant studies published in journals, conference proceedings, and technical reports. We used a systematic approach, including the detailed inclusion/exclusion criteria given in Table~\ref{tab:IncExc} to select the relevant primary studies. Initially both authors applied the study selection process to a subset of the results returned by the key word based search. After verifying that both authors selected the same set of studies, the first author applied the study selection process to the rest of the results returned by the key word based search.

In addition we examined the reference lists of the selected primary studies to identify any additional studies that relate to our search focus. We found 13 additional studies related to our search focus. These studies were found by the key word based search, but did not pass the title based filtering. This indicates that selecting studies based on the title alone may not be reliable and to improve the reliability we might have to review the abstract, key words and conclusions before excluding them. This process would be time consuming due to the large number of results returned by the key word based search. After selecting the primary studies, we used data extraction forms to extract the relevant information consistently while reducing bias. Extracted information was validated by both authors.

We used the quality assessment questions given in Table~\ref{tab:qualityForQuant} and Table~\ref{tab:qualityForQual} for assessing the quality of the selected primary studies. All selected primary studies are of high quality. The primary studies are a mixture of observational and experimental studies.

One weakness is the reliance on the key word based search facilities provided by the three databases for selecting the initial set of papers. We cannot ensure that the search facilities return all relevant studies. But, the search process independently returned all the studies that we previously knew as relevant to our research questions.

Many primary studies were published in venues that are not related to software engineering. Therefore, there may be solutions provided by the software engineering community for some of the challenges presented in Section~\ref{sec:challenges} such as oracle problems. But we did not find evidence of wide use of such solutions by the scientific software developers.  

\subsection{Contribution to research and practice community}

To our knowledge, this is the first systematic literature review conducted to identify software testing challenges, proposed solutions, and unsolved problems in scientific software testing. We identified challenges in testing scientific software using a large number of studies. We outlined the solutions used by the practitioners to overcome those challenges as well as unique solutions that were proposed to overcome specific problems. In addition we identified several unsolved problems.

Our work may contribute to focusing research efforts aiming at the improvement of testing of scientific software. This SLR will help the scientists who are developing software to identify specific testing challenges and potential solutions to overcome those challenges. In addition scientist developers can become aware of their cultural differences with the software engineering community that can impact software testing. Information provided here will help scientific software developers to carry out systematic testing and thereby improve the quality of scientific software. Further, there are many opportunities for software engineering research to find solutions to solve the challenges identified by this systematic literature review.

\section{Conclusion and future work}
\label{sec:conc}

Conducting testing to identify faults in the code is an important task in scientific software development that has received little attention. In this paper we present the results of a systematic literature review that identifies specific challenges faced when testing scientific software, how the challenges have been met and any unsolved challenges.. Below we summarize the answers to our four research questions:

\textbf{RQ1: How is scientific software defined in literature?} Scientific software is defined as software with a large computational component. Further, scientific software is usually developed by multidisciplinary teams made up of scientists and software developers. 

\textbf{RQ2: Are there special characteristics or faults in scientific software or its development that make testing difficult?} We identified two categories of challenges in scientific software testing: (1) testing challenges that occur due to characteristics of scientific software such as oracle problems and (2) testing challenges that occur due to cultural differences between scientists and the software engineering community such as viewing the code and model as inseparable entities.

\textbf{RQ3: Can we use existing testing methods (or adapt them) to test scientific software effectively?} A number of studies report on testing at different levels of abstraction such as unit testing, system testing and integration testing in scientific software development. Few studies report the use of unit testing frameworks. Many studies report the use of a pseudo oracle or experimental results to alleviate the lack  of an oracle. In addition, several case studies report using metamorphic testing to test programs that do not have oracles. Several studies developed techniques to overcome challenges in test case creation. These techniques include the combination of randomly generated test cases with specially designed test cases, generating test cases by considering dependencies among input parameters, and using symbolic execution to generate test data for floating point programs. Only two studies use test coverage information.

\textbf{RQ4: Are there challenges that could not be met by existing techniques?} Oracle problems are prevalent and need further attention. 

Scientific software poses special challenges for testing. Some of these challenges can be overcome by applying testing techniques commonly used by software engineers. Scientist developers need to incorporate these testing techniques into their software development process. Some of the challenges are unique due to characteristics of scientific software, such as oracle problems. Software engineers need to consider these special challenges when developing testing techniques for scientific software. 
\section{Acknowledgments}
This project is supported by Award Number 1R01GM096192 from the
National Institute of General Medical Sciences. The content is solely the
responsibility of the authors and does not necessarily represent the official
views of the National Institute of General Medical Sciences or the National
Institutes of Health. We thank the reviewers for their insightful comments on earlier versions of this paper.


\bibliographystyle{elsarticle-harv}

\bibliography{./refs/MetamorphicTesting,./refs/Misc,./refs/SysReview_primaryPapers,./refs/PapersAboutSysLitSurveys,./refs/myPubs,./refs/SLRs}


\end{document}